\documentstyle[11pt,newpasp,twoside]{article}
\def\edcomment#1{\iffalse\marginpar{\raggedright\sl#1\/}\else\relax\fi}
\reversemarginpar

\begin{document}
\title{
3D Identification and  Reconstruction of z $\sim$ 1 Clusters: 
Prospects for the DEEP2 Redshift Survey}
 \author{C. Marinoni, M. Davis, J. Newman, A. Coil}
\affil{University of California at Berkeley, Berkeley, CA, 94720-3411, USA}

\begin{abstract}

We have developed a geometrical method based on 3D Voronoi polyhedra
and Delaunay tessellation for identifying and reconstructing clusters
of galaxies in the next generation of deep, flux-limited redshift
surveys.  We here describe this algorithm and tests of it using mock
catalogs that simulate the DEEP2/DEIMOS redshift survey, 
which will begin observations in the Spring of 2002 and will
provide a detailed three dimensional map of the large scale structure
up to redshift 1.5.
\end{abstract}

\section{Introduction}
Over the past few years there has been considerable work on the
detection of two-dimensional galaxy overdensities in wide field
imaging surveys for subsequent spectroscopic follow up.  However, these
systematic searches for clusters, fuelled by the availability of large
CCD camera mosaics, are in many cases biased towards high density peaks 
of the galaxy distribution.

With the next generation of large multi-object spectrographs in their
final construction phase, deep redshift surveys (Keck/DEEP2 and
VLT/VIRMOS) are well within the horizon.   It is now worthwhile to address the more
ambitious task of constructing a three dimensional, statistically
complete sample of high redshift galaxy clusters.  Here we describe a
fully automated and objective  algorithm for identifying and
reconstructing galaxy systems, based upon 3D Voronoi polyhedra and
Delaunay triangulation.

\section{Method}
A Voronoi partition of the space into minimally sized convex polytopes
is a natural way to measure packing. A Voronoi polyhedron is the
uniquely defined  region of space around a galaxy (seed),
within which each point  is closer to the 
seed than to any other galaxy.  The faces of the Voronoi cell
are formed by planes perpendicular to the vectors between a galaxy and
its neighbors.  The volume inside the polyhedron is inversely
proportional to the packing efficiency of its central galaxy;
a large cell volume is indicative of an isolated galaxy.
The Delaunay triangulation is the geometrical dual of the Voronoi
partition and is defined by the tetrahedron whose
vertices are the 4 galaxies  with the property that
the uniquely determined circumscribing sphere does not contain any
other galaxy.

We thus may identify high-density peaks by selecting as ``cluster
seeds'' the centers of all the Voronoi cells with volume smaller than
some threshold, and then use the scale length of
the Delaunay mesh to infer the strength of the physical aggregation.
We then can define an adaptive cylindrical window in redshift space (elongated
in the $z$ direction) with dimensions determined by the local scale factor
and  process all the Delaunay connected galaxies 
with a rapidly converging  ``inclusion-exclusion'' logic  
to identify cluster members. 

In our algorithm, the parameters
defining the adaptive search window are only weakly dependent on
the galaxy density gradient which inevitably occurs in a flux limited
survey, since our smoothing procedure immediately identifies
regions of enhanced clustering where galaxies are generally more luminous.
Moreover  the number of cluster interlopers aggregated into a
system in a redshift space analysis is minimized and we
simultaneously obtain a non-parametric local
estimate of the surrounding density environment for each galaxy and a
quantitative measure of the distribution of cosmological voids in the
survey volume.

\begin{figure}
\plotfiddle{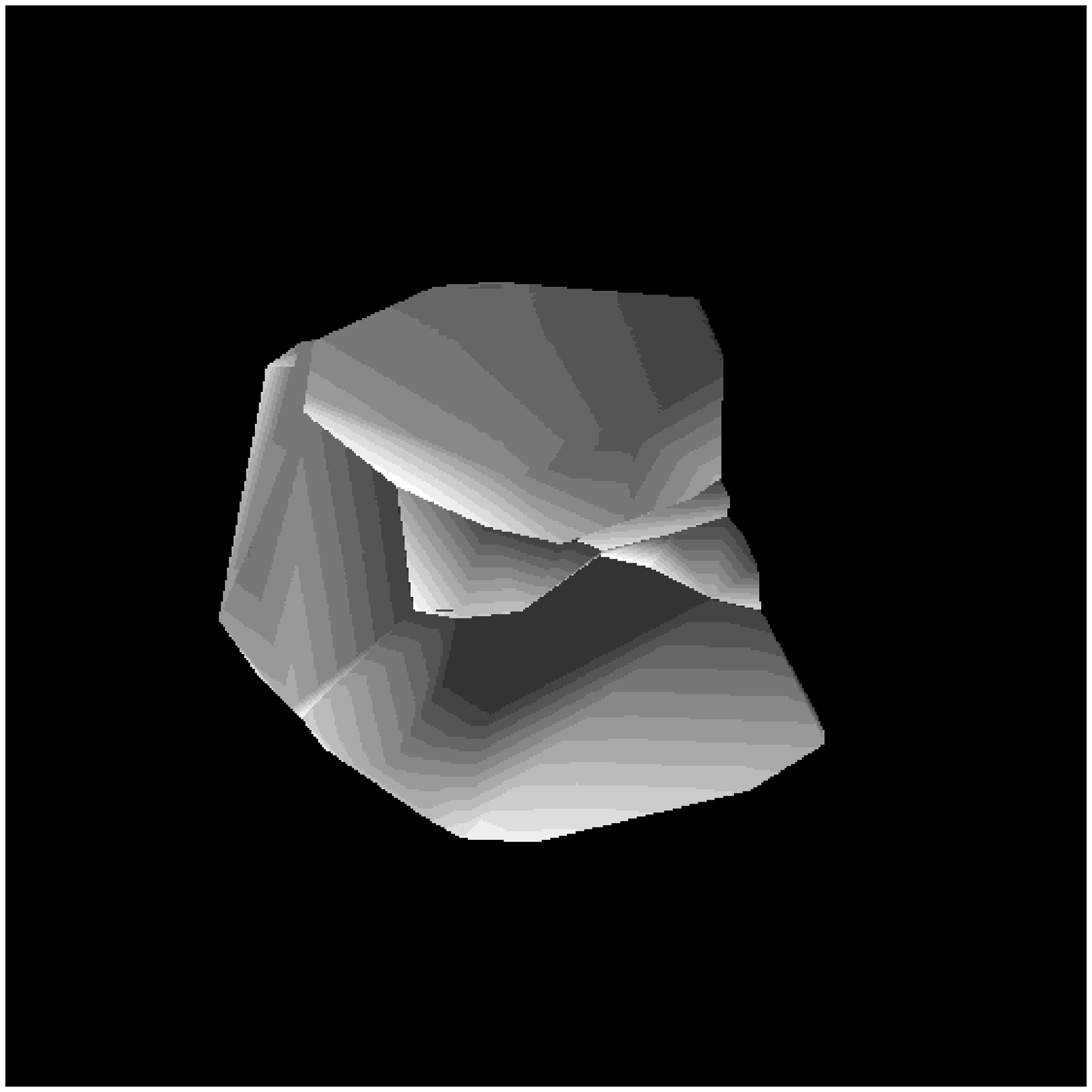}{4.5cm}{0}{35}{35}{-200}{-80}
\plotfiddle{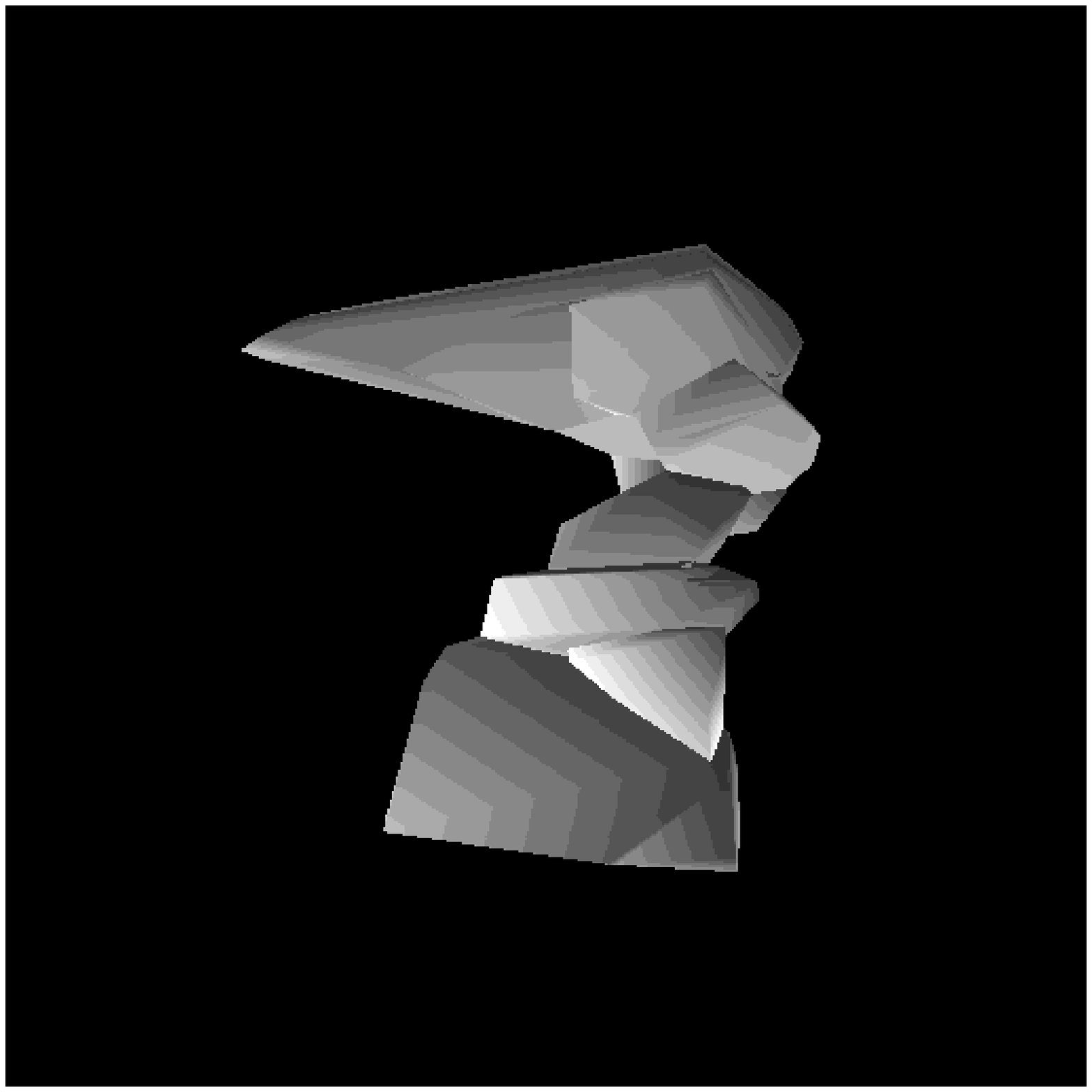}{0cm}{0}{35}{35}{-20}{-57}
\caption{
Voronoi reconstruction of a cluster with 10 galaxies in the DEEP2 mock catalog.
 The Voronoi cells encompassing the  cluster are shown in real space ({\em left})
and in redshift space ({\em right}).}
\end{figure}

\section{Results}
We have tested our reconstruction scheme using mock catalogs modelling
the DEEP2 Redshift Survey (Davis et al. 2000) derived from GIF
simulations (Coil et al. 2001). This
survey will obtain high quality spectra for $\sim$ 60000 galaxies
between z=0.7-1.5 using the Keck 2 Telescope and DEIMOS spectrograph,
with the twin goals of studying the evolution of the properties and
the large-scale clustering of high redshift galaxies.  We used a LCDM
model with $\Omega_m =0.3$, $\Omega_{\Lambda}=0.7$, $h=0.7$ and
$\sigma_8 =0.9$.  In order to mimic the selection function of the
magnitude limited survey we make a rest-frame B-band cut of 23.4,
which at z $\sim$ 1 corresponds to an apparent I-band
magnitude with the appropriate K-correction matching the DEEP2
photometric selection criteria ($I_{AB}<23.5$).  In Fig. 2a we present
a mock catalog for one DEEP2-like field collapsed 
along the smallest axis, covering the redshift range
$z$=0.7-1.2 (a depth which corresponds to selecting the 1200 l/mm
grating in the DEIMOS spectrograph) and containing a volume of roughly $10^6$ Mpc
h$^{-3}$ ($\sim$ 15000 galaxies).  We have used six independent
mock catalogs to determine how well the algorithm performs in
identifying clusters of galaxies. 
                   
To determine the ``true'' distribution of clusters, we have applied in
real space (volume-limited simulation) a standard percolation algorithm optimised for selecting
virialized objects (mean overdensity $\sim$ 180).  The resulting
distribution of those clusters with more than 5 members is shown in
Fig 2b.  Cluster candidates selected by the Voronoi--Delauney algorithm applied
in redshift space (flux-limited simulation) are shown in Fig 2c. 
Note how the large-scale
pattern defined by galaxy systems and the associations of clusters in
higher order structures reproduce in an unbiased way the underlying
real space landscape.

\begin{figure}
\plotfiddle{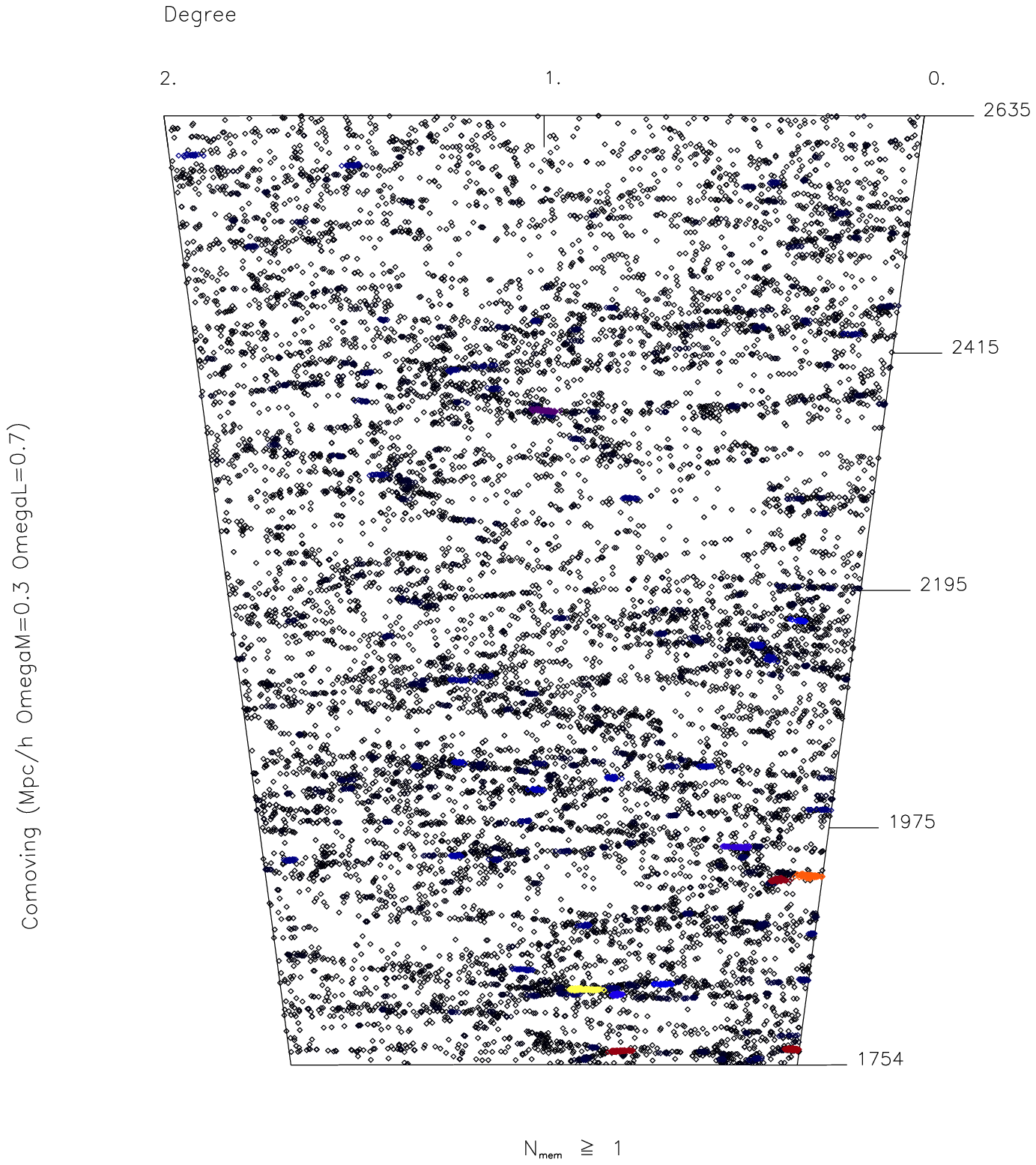}{6.5cm}{0}{35}{55}{-250}{-110}
\plotfiddle{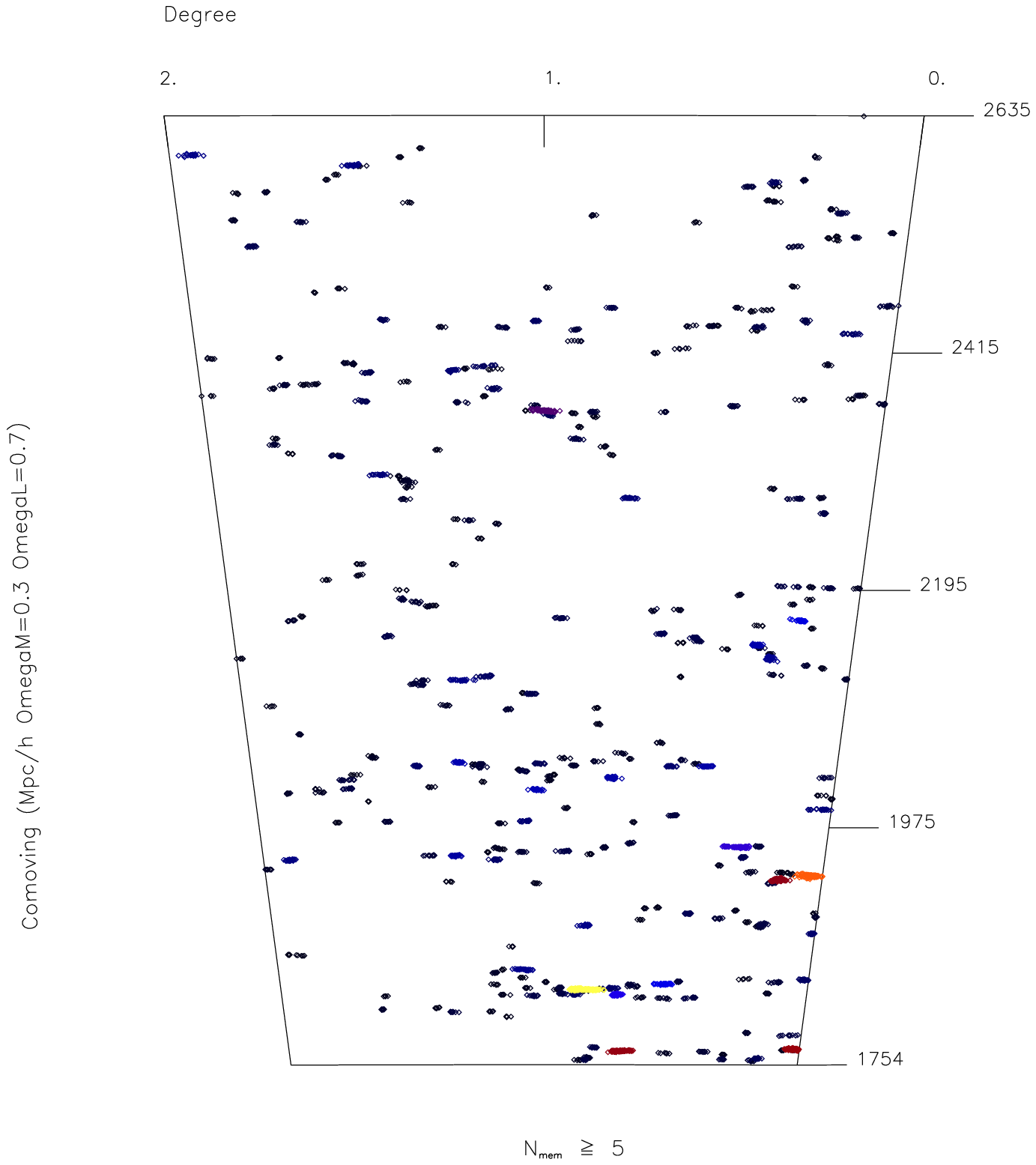}{0cm}{0}{35}{55}{-100}{-85}
\plotfiddle{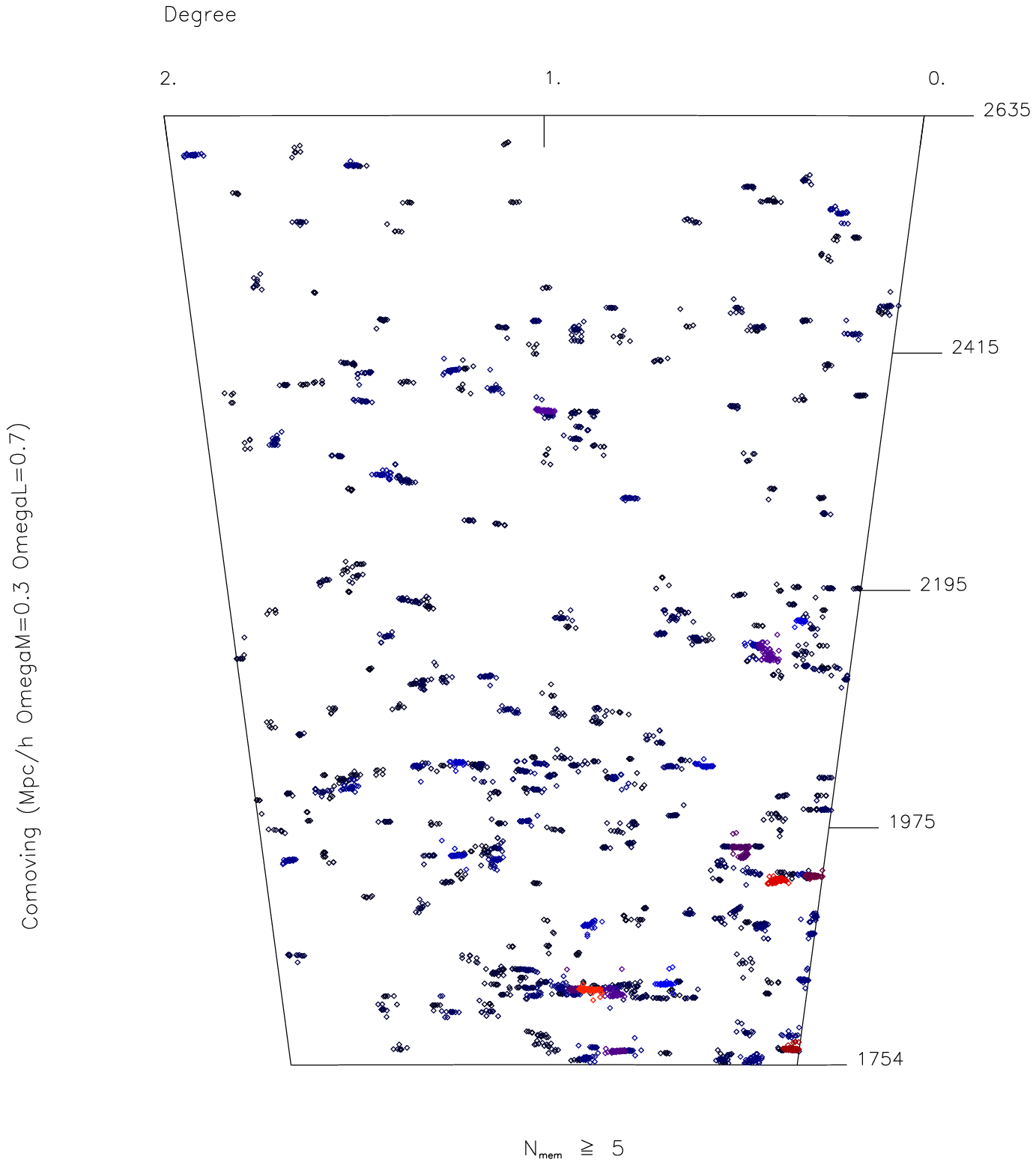}{0cm}{0}{35}{55}{+55}{-60}
\caption{{\em Left:} One DEEP2 mock catalog in the redshift range z=(0.7,1.2).
{\em Center:} large-scale spatial distribution of clusters with more that 5 members 
identified in real space using a ``Friends of Friends'' algorithm.
{\em Right:} Voronoi-Delaunay reconstruction in redshift space.} 
\end{figure}

 
\begin{figure}
\plotfiddle{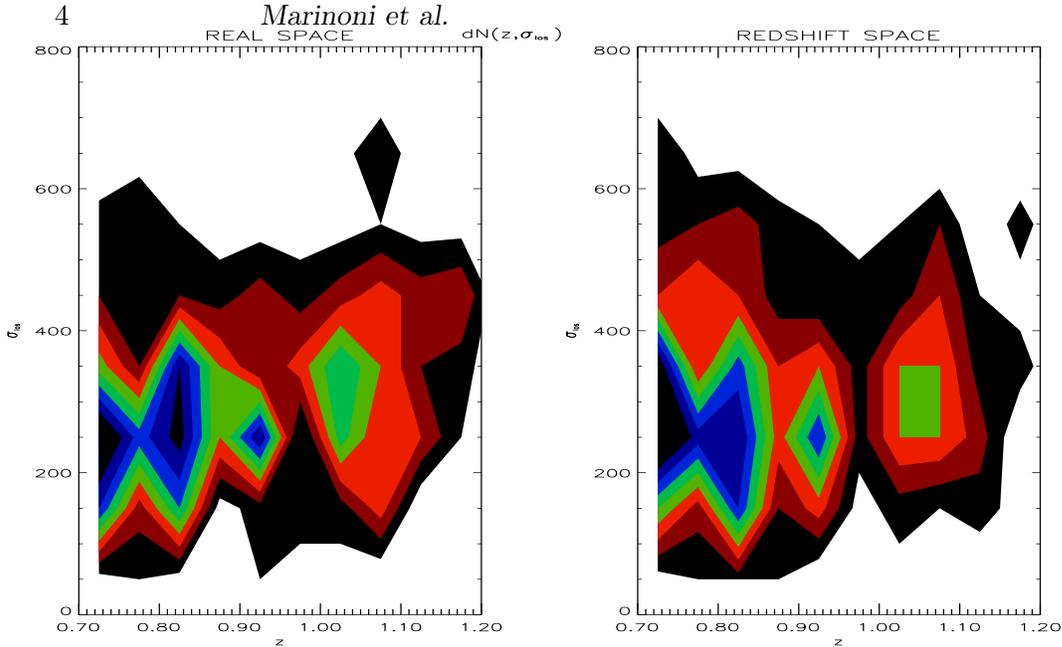}{7.cm}{0}{90}{51}{-290}{-93}
\caption{Differential number count of cluster as a function 
of the redshift z and the projected velocity dispersion $\sigma$.
The parent distribution of
cluster counts identified in real space is shown on the left,
while the
observed distribution recovered in redshift space with our method 
is plotted on the right. Density levels, smoothed by 0.05 in z and 50 km/s
in $\sigma$, are spaced with $\Delta$=2 intervals.}
\end{figure}

\section{Statistical tests}
The evolution of the comoving abundance of clusters as a function of their 
velocity dispersion $\sigma$ and redshift $z$  is a sensitive function of
cosmological parameters (see Newman et al. 2001).   
The statistical significance and robustness of the cluster abundance test depends 
critically  on an unbiased mapping of the distribution properties of the cluster
observables between real to redshift space. 

We have investigated the differential and integral distribution functions
of the number of reconstructed clusters as a function of
their richness, velocity dispersion and redshift.   
The Kolmogorov-Smirnov statistical test confirms the 
similarity between the real and reconstructed distributions, and thus the
reliability of our algorithm. This can be seen visually  in Fig. 4 where
we show the two-dimensional distributions of the real and reconstructed  
clusters in  redshift z and  projected velocity dispersion $\sigma$.

\bigskip

The DEEP2 survey is a collaborative project among astronomers
at UC, Caltech and the Univ. of Hawaii. Details of the project
can be found at the URL http://astro/berkeley.edu/deep.

\end{document}